# Ultrafast Non-Volatile Weyl LuminoMem for Mid-Infrared In-Memory Computing

Delang Liang[1,2], Shiyu Wang[1], Yan Wang[3], Dong Li[2], Yuchun Chen[1], Bin Cheng[4], Mingyang Qin[1], Dehong Yang[1], Jie Sheng[1], Lin Li[4], Changgan Zeng[4], Dong Sun[1,5,6,7,8]*, Anlian Pan[2,9]* and Jing Liu[3]*

[1]Center for Quantum Materials, School of Physics, Peking University, Beijing, China.
[2]Key Laboratory for Micro-Nano Physics and Technology of Hunan Province, Hunan Institute of Optoelectronic Integration, College of Materials Science and Engineering, Hunan University, Changsha, China.
[3]State Key Laboratory of Precision Measurement and Instrumentation, School of Precision Instruments and Opto-electronics Engineering, Tianjin University, Tianjin, China.
[4]CAS Key Laboratory of Strongly Coupled Quantum Matter Physics, and Department of Physics, University of Science and Technology of China, Hefei, Anhui, China.
[5]School of Physics and Laboratory of Zhongyuan Light, Zhengzhou University, Zhengzhou, China.
[6]Collaborative Innovation Center of Quantum Matter, Beijing, China.
[7]Frontiers Science Center for Nano-optoelectronics, School of Physics, Peking University, Beijing, China.
[8]Beijing Key Laboratory of Quantum Devices, Peking University, Beijing, China
[9]School of Physics and Electronics, Hunan Normal University, Changsha, China.

*Email: Jingliu_1112@tju.edu.cn; anlian.pan@hnu.edu.cn; sundong@pku.edu.cn;

**Abstract**

Integrated optoelectronic systems strive to combine the logic/memory density of electronics with the bandwidth of photonics, but monolithic realization is impeded by the inefficient electronic-to-photonic interface. Current architectures rely on separate readout circuitry and modulators, creating bottlenecks in energy and latency, while existing direct transduction methods often compromise on switching speed or non-volatility. Here, we report an ultrafast, non-volatile optoelectronic memory, named LuminoMem, that integrates electrical storage and mid-infrared light emission in a single device. The device utilizes a floating-gate architecture, in which the Weyl semiconductor tellurium serves simultaneously as a charge-trapping storage layer and an emissive medium. This design enables nanosecond-scale electrical programming of non-volatile photoluminescence at 3.4 μm, allowing direct optical access to stored states without external modulation. We demonstrate 4-bit (16-level) optical storage capacity and validate the device's performance through neural network simulations that achieve high accuracy on the Fashion-MNIST dataset. By effectively bridging the gap between electronic storage and mid-infrared photonics, the demonstrated mid-infrared LuminoMem provides a hardware foundation for promoting current computation efficiency and potential intelligent platforms that co-integrate computing, memory, and sensing capabilities.

**Introduction**

The quest for next-generation intelligent platforms is accelerating the development of integrated optoelectronic systems that combine the strengths of electronics for dense, low-noise logic and memory with the strengths of photonics for ultrafast, massively parallel data transport and processing[1-4]. By bringing these capabilities together on chip, such architectures could move beyond the performance and energy-efficiency ceilings of purely electronic hardware, and enable hardware-level co-integration of computing with communication and sensing[5-7]. However, monolithic realization remains elusive, limited by a central bottleneck: inefficient and power-hungry transfer of information across the electronic-to-photonic interface[4,8-10].

A primary source of inefficiency lies in converting electrically stored data in electronic memory into optical signals. In conventional architectures, data are first retrieved from memory by dedicated electronic readout circuitry and then routed to a separate optical modulator for electro-optic conversion (Fig. 1a), a two-step workflow that incurs substantial energy overheads and added latency[8,11]. This architectural discontinuity erodes the intrinsic advantages of photonics and frustrates truly dense integration, since the material stacks (e.g., LiNbO$_3$, III–V, phase-change or plasmonic platforms) and the corresponding fabrication process flows that enable high-performance modulators are frequently incompatible with high-volume on-chip integrated manufacturing[12-16].

Attempts to achieve direct electro-optic transduction on a single chip have introduced new tradeoffs. For example, doping based approaches regulate optical emission or transmission by electrically tuning the carrier concentration, but such effects lack non-volatility[17-21]. Phase change materials (PCMs) offer an appealing route to non-volatile photonic modulation through electrically driven phase transitions, but their switching speeds are often too slow to match the throughput in high-bandwidth photonic systems[14,22-26]. Ferroelectric platforms can, in principle, enable non-volatile functionality through polarization switching, but they remain difficult to scale because their materials and integration flows are often incompatible with standard complementary metal-oxide-semiconductor (CMOS) manufacturing[27-30].

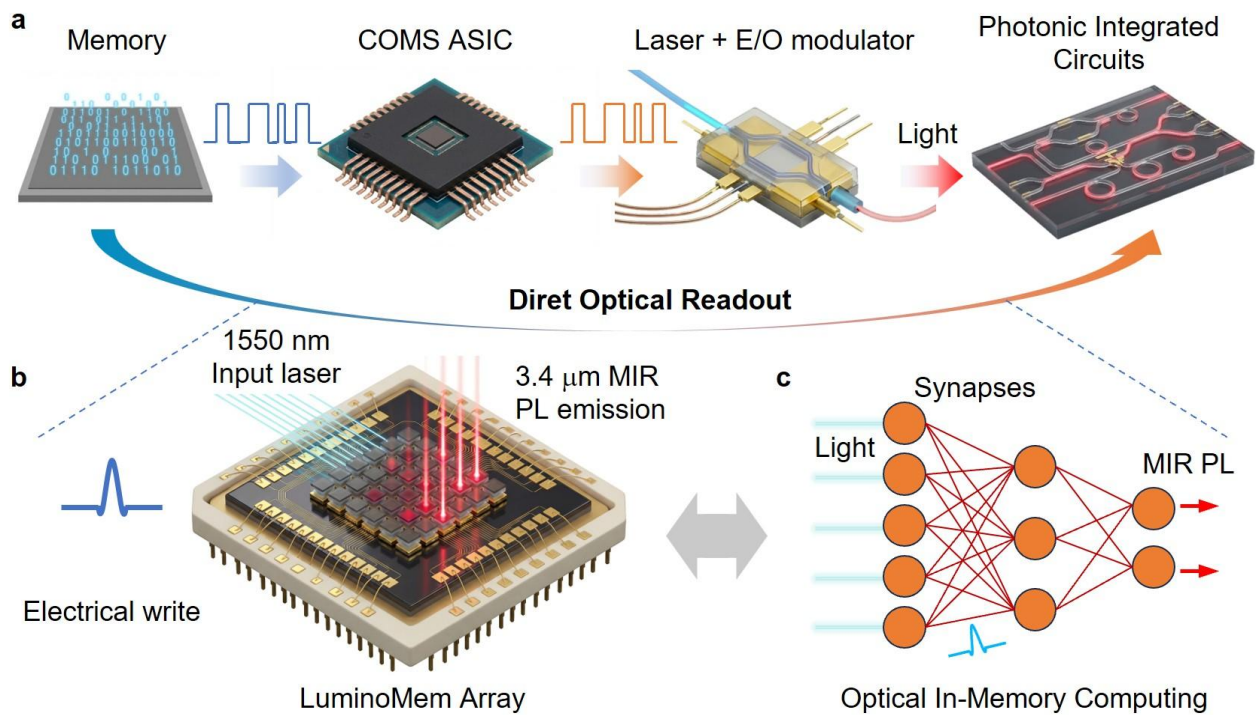

**Fig. 1 | Concept and mechanism of LuminoMem. a**, Conventional hardware bottleneck and the redundant data chain. Schematic of the typical electronic-to-photonic interface where data must be fetched from memory, processed by a CMOS application-specific integrated circuit (ASIC), and converted into the optical domain via an external E/O modulator before entering the photonic integrated circuits for further processing. **b**, Illustration of the proposed solution featuring a LuminoMem array where MIR PL is emitted directly upon electrical programming. **c**, Monolithic architecture for in-memory computing.

To address these challenges, we introduce LuminoMem, an ultrafast and non-volatile optoelectronic memory that monolithically co-integrates memory and light-emitting functionality based on photoluminescence (PL) from a semiconductor. The LuminoMem demonstrated in this work uses a floating-gate architecture in which the Weyl semiconductor tellurium (Te) serves both as the charge-trapping layer and as the light emissive medium. This design enables ultrafast (nanosecond-scale) electrical programming that precisely and persistently tunes the PL intensity. By eliminating separate readout circuitry and external optical modulators, LuminoMem enables direct optical readout of non-volatile electrical states (Fig. 1b). In addition, Te-LuminoMem operates in the MIR region at ~3.4 μm, leveraging the narrow bandgap (~365 meV) of Te. This wavelength lies in a particularly important spectral region with strong molecular "fingerprints"[31] and favorable atmospheric transmission windows[32], creating opportunities in gas sensing[33], environmental monitoring[34], industrial process control[35], and security/homeland-defense applications—capabilities that are far less accessible in the more commonly explored visible-to–near-infrared range where most prior devices operate[27,30].

The LuminoMem demonstrated in this work exhibits robust multi-level memory characteristics, achieving 4-bit (16-level) precision in its optical readout. We further demonstrate its potential for in-memory computing by simulating an array of LuminoMem devices within an artificial neural network, successfully performing image classification tasks on the Fashion-MNIST dataset. This work overcomes key bottlenecks by strategically leveraging the advantages of both electrons and photons, resulting in a high-speed non-volatile memory-light emitter. This performance breakthrough paves the way for greatly enhancing the efficiency of current computation hardware and the on-chip integration of memory, computing and sensing capabilities (Fig. 1c).

## Results and Discussions

### Monolithic device design and basic characterization

The LuminoMem device is built on a vertically stacked floating-gate architecture that combines two-dimensional (2D) materials with the recently rediscovered Weyl semiconductor Te. As shown in Fig. 2a, the device comprises a graphite/hexagonal boron nitride/Te (Gr/h-BN/Te) van der Waals (vdW) heterostructure fabricated on a $SiO_2$/Si substrate. In this configuration, the Te nanoflake simultaneously functions as the floating charge-storage layer and the light-emitting medium. Te has been identified as a Weyl semiconductor with a nearly direct bandgap of ~365 meV, corresponding to MIR emission centered at 3.4 μm. Our recent work shows that the PL of Te can be continuously tuned by its doping level[36]; consequently, changes in the charge-storage state directly modulate the emission intensity, enabling electrical programming with a direct MIR optical readout. In the device stack, an h-BN nanoflake acts as the tunneling barrier, electrically isolating the Te floating layer from the few-layer graphite top electrode, which is held at ground potential. The p-doped Si substrate serves as the control gate ($V_{CG}$), allowing modulation of the channel potential and the stored charge state.

Figure 2b presents an optical micrograph of the LuminoMem device, with dashed lines outlining the key components: few-layer graphite (yellow), h-BN (blue), and Te nanoflake (red).

The Gr/h-BN heterostructure was assembled using a polymer-free dry transfer method and precisely aligned atop the Te nanoflake, as described in the Methods section. The high-quality and atomically flat interfaces of these layered materials are critical for efficient charge tunneling and reduced scattering[37]. Polarization-dependent Raman spectroscopy confirmed the high crystalline quality and orientation of the Te nanoflake, with signature $E_1$-TO (92 cm$^{-1}$), $A_1$ (121 cm$^{-1}$), and $E_2$ (143 cm$^{-1}$) modes exhibiting maximum intensity for polarization parallel to the c-axis (Fig. 2c)[38].

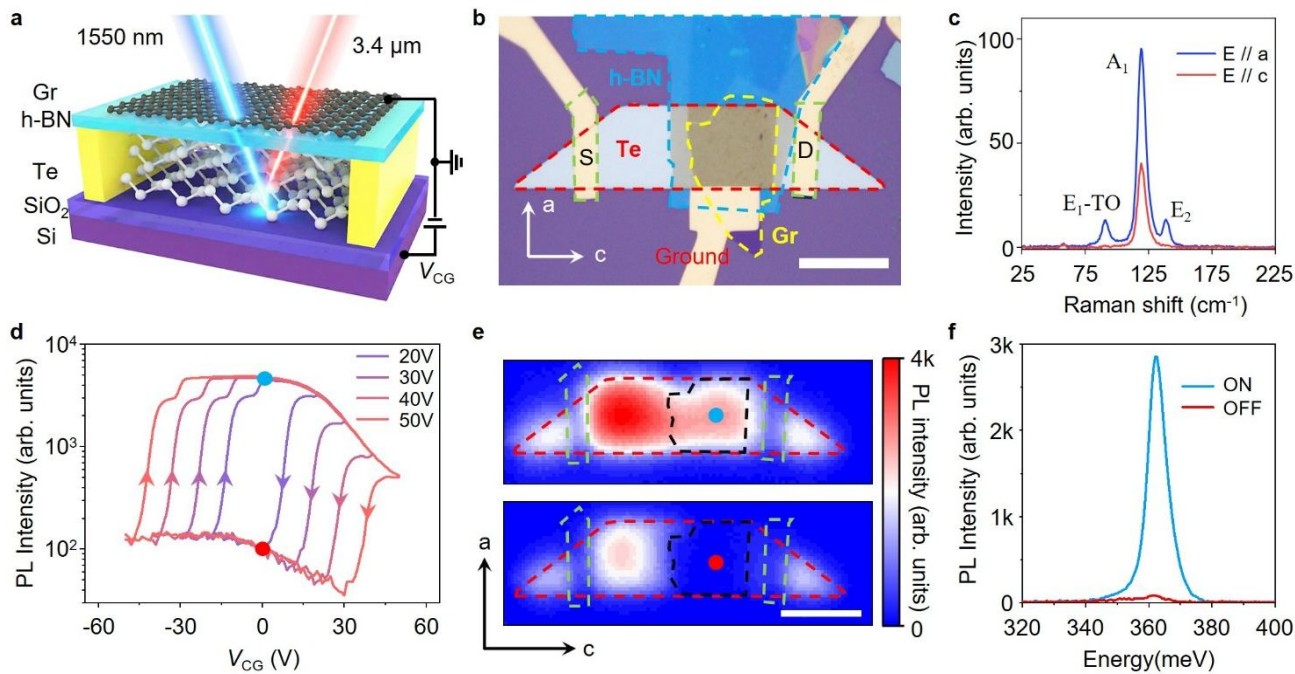

**Fig. 2 | Device characterization. a,** Schematic of the Gr/h-BN/Te vdW heterostructure memory device on a SiO$_2$/Si substrate. The doped Si serves as the control gate, and the top graphite electrode is grounded. **b,** Optical micrograph of a typical floating-gate device. Dashed lines outline the Te nanoflake (red), graphite floating gate (yellow), and h-BN flake (blue). Scale bar, 30 μm. **c,** Polarization-dependent Raman spectroscopy of Te nanoflakes using 633-nm laser excitation with polarization oriented either perpendicular or parallel to the c-axis. **d,** MIR PL intensity versus control-gate voltage ($V_{CG}$) measured with $V_{CG}$ dual-sweep ranging from ±20 V to ±50 V in 10 V steps. The prominent hysteresis demonstrates programmable charge trapping. **e,** Spatially mapped MIR PL intensity across the entire device after applying $V_{CG}$ =-40 V (top panel) and +40 V (bottom panel). Black dashed lines mark the junction area. Scale bar is 30 μm. **f,** MIR PL spectra corresponding to the 'ON' and 'OFF' states at the positions marked by the red and blue points in **e**, respectively.

**Non-volatile MIR PL modulation**

Next, we study the non-volatile hysteresis character of the LuminoMem device. To readout the memory state, we optically excite the PL signal of the device, which reflects the charge storage level of Te according to our previous gate dependent PL studies[36]. Owing to the narrow bandgap of Te, a low-energy telecommunication-band laser at 1550 nm is used as the optical read input as its photon energy is insufficient to surmount the high internal energy barrier of h-BN, thereby enabling non-destructive optical readout. Figure 2d presents the MIR PL intensity variation of the device as a function of the control-gate voltage ($V_{CG}$) obtained at 25 K, by sweeping the $V_{CG}$ bidirectionally between the negative and positive limits. The PL curves show prominent hysteresis, in which the PL intensity at zero gate bias depends strongly on the prior gate voltage history, resembling the transfer characteristics of electrical floating-gate memories.

Specifically, applying a large positive gate bias depletes carriers in the Te nanoflake (program operation), driving the device into a low-PL-intensity ("OFF") state. This low-PL state persists as the gate voltage is swept back to 0 V. When the gate voltage is subsequently swept from 0 V toward negative values, the Te nanoflake remains depleted until a threshold negative bias is reached, at which point the gate field is strong enough to drive electrons back to the graphite electrode and restore the hole concentration in the Te nanoflake (erase operation), resulting in a high-PL-intensity ("ON") state. As the gate voltage is then swept from this large negative bias back to 0 V, the carriers are largely retained in the Te channel, so the device remains in the high-PL state. The memory window (ΔV), defined as the difference in the threshold voltage between forward and backward sweeps, reflects the charge storage capacity of the device. A larger memory window indicates a greater number of charges stored in the Te floating gate, which enhances the reliability of the memory. As the gate-voltage sweep span expanded from 40 V (±20 V) to 100 V (±50 V), the memory window increased from 20 V to 80 V, comparable to that of conventional electrical memories.

To confirm that this hysteresis originates from charge storage in the floating Te layer rather than interface traps, we investigated the device performance in a different configuration, in which the Te nanoflake, instead of the graphite layer, is electrically grounded as shown in Supplementary Fig. 1. In this configuration, the PL intensity shows negligible hysteresis, which is obviously different from the performance when the Te layer is floated, confirming non-volatile charge trapping in the Te floating layer. Furthermore, the reproducibility of this non-volatile Weyl LuminoMem was verified across six additional devices. As illustrated in Supplementary Fig. 2, all devices consistently exhibit a clear and programmable memory window.

Figure 2e presents spatial PL mapping of the device to visualize the non-volatile switching behavior across the entire device. When a negative gate voltage ($V_{CG}$ = -40 V) is applied, the Te nanoflake is set to a high-emission "ON" state (see Fig. 2e top panel). The region beneath the graphite gate (outlined by the black dashed line) exhibits a slightly lower PL intensity than the exposed Te area, due to optical absorption by graphite. After applying a positive gate voltage ($V_{CG}$ = +40 V), the device switches to a uniform low-emission "OFF" state across the entire active region (see Fig. 2e bottom panel). This further confirms efficient charge carrier localization in the floating-gate region. Spectral measurements (Fig. 2f) reveal that switching between the "ON" and "OFF" states produces substantial PL intensity modulation while maintaining a constant peak wavelength. This wavelength stability is critical for maintaining signal purity and preventing chromatic dispersion, ensuring crosstalk-free operation in MIR integrated photonic systems. To evaluate the design efficacy of this architecture, we further compared our primary Gr/h-BN/Te structure with an alternative Te/h-BN/Gr configuration, in which the Te nanoflake serves as the grounded channel and graphite acts as the floating gate (Supplementary Fig. 4). The results show that the direct injection of carriers into the Te emitter in the Gr/h-BN/Te design facilitates more efficient control over radiative recombination than remote field-effect modulation in the alternative geometry, leading to a significantly enhanced optical on/off ratio.

**Electro-optical conversion mechanism**

Figure 3a illustrates the operating mechanism of the non-volatile LuminoMem. The device operation relies on the controlled modulation of the majority carrier (hole) density in the Te nanoflake, which serves simultaneously as the charge storage medium and the photon emitter. During the program operation (Fig. 3a(i)), a positive gate voltage ($V_{CG} > 0$) induces a steep band bending across the h-BN tunneling barrier, facilitating electrons to tunnel from the graphite control gate into the Te floating layer via the Fowler-Nordheim mechanism. The injected electrons recombine with the majority holes in the Te layer, leading to the reduction of carriers in the Te layer. This reduction in carrier doping density significantly decreases the radiative recombination rate, resulting in a low PL intensity, or the "OFF" state. After the voltage is removed, the injected electrons stay in the Te layer due to the h-BN barrier, and thus, the device remains in the "OFF" state (Fig. 3a(ii)). Conversely, during the Erase operation (Fig. 3a(iii)), a negative gate voltage pulse ($V_{CG} < 0$) causes electrons to tunnel out of the Te layer back into the gate electrode, which is equivalent to hole accumulation in the Te layer. This high doping concentration of majority carriers boosts the radiative recombination efficiency, switching the device to a high PL intensity, or the "ON" state. When the negative gate voltage is removed (Fig. 3a(iv)), the high hole doping state remains in the Te layer since few electrons can tunnel into the Te layer through the h-BN barrier. This electrically controlled, reversible charge tunneling mechanism allows for precise and non-volatile modulation of the PL emission intensity.

Another key feature of this architecture is the non-destructive optical readout. The photon energy of the 1550 nm readout laser (~ 0.8 eV) is insufficient to excite the stored electrons in the Te layer over the high energy barrier of h-BN back to the graphite gate, thus preserving the charge state during measurement. This mechanism is validated by contrasting the device's response to short-wavelength visible light (520 nm, ~2.38 eV) presented in Supplementary Fig. 4. Unlike the non-perturbing NIR excitation, the higher energy at 520 nm photons enables stored carriers to surmont the h-BN barrier and return to the graphite gate. This pronounced wavelength selectivity, which is stable under NIR probing yet erasable under visible illumination, provides direct evidence that charge confinement is governed by the energy offsets of the vdW heterostructure, enabling all-optical erasure via photo-induced carrier de-trapping[39-42].

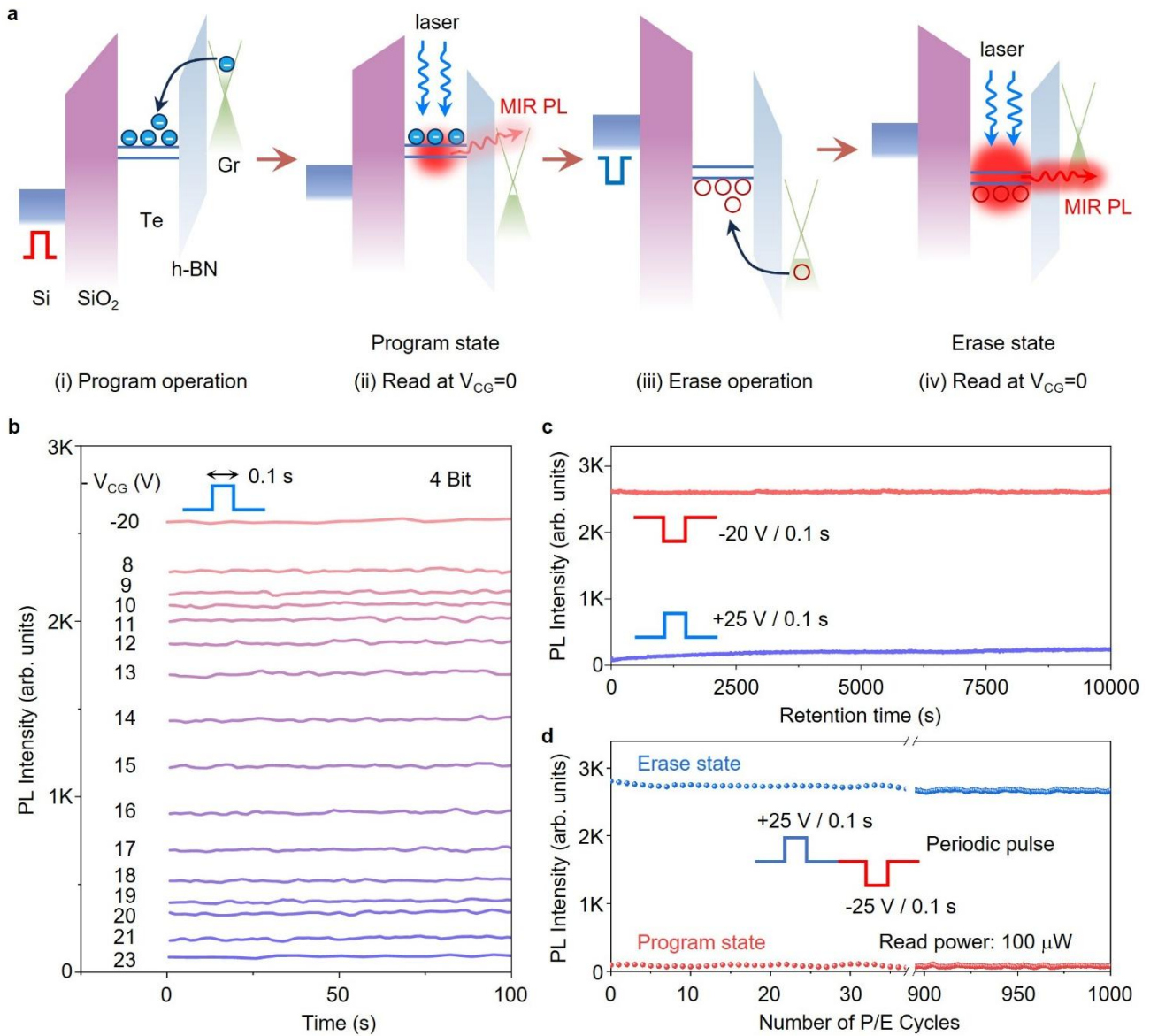

**Fig. 3 | Operating mechanism and performance benchmarks. a,** Schematic energy-band diagrams illustrating the device's operating principles during program, read and erase operations. The red glow intensity represents the corresponding radiative recombination/PL signal. **b,** Demonstration of 16 distinct, non-volatile memory states (4-bit storage) achieved by modulating the programming pulse amplitude from -20 V to 23 V (pulse duration: 0.1 s). **c,** Data retention tests for the binary memory states. The programmed and erased states were set using +25 V and -20 V 0.1 s pulses, respectively, and show negligible degradation in PL intensity over $10^4$ s. **d,** Endurance cycling test, showing robust switching between the programmed and erased states for over 1,000 cycles (pulse conditions: ± 25 V, 0.1 s duration).

**Multi-level storage and reliability characterization**

Beyond binary operation, the analog nature of the charge accumulation in the floating gate structure enables multi-level data storage, a critical feature for high-density memory and neuromorphic computing applications. By fine-tuning the amplitude of the programming voltage pulses, we can precisely control the quantity of charges injected into the Te floating layer, thereby achieving fine-grained modulation of the PL readout intensity. Figure 3b demonstrates the realization of 4-bit (16 distinct levels) memory capacity. This was achieved by applying programming pulses with amplitudes ranging from -20 V to +23 V (pulse duration: 0.1 s), which produced clearly separated and stable PL intensity levels. The well-defined separation between adjacent levels confirms the deterministic and highly controllable nature of the electro-optical modulation.

The device's reliability was rigorously evaluated through data retention and endurance tests. As shown in Fig. 3c, the programmed (+25 V, "OFF") and erased (-20 V, "ON") states exhibit exceptional stability, with negligible degradation in PL intensity over $10^4$ s. This outstanding data retention underscores the excellent charge confinement provided by the h-BN/Te/SiO$_2$ heterostructure[43,44]. Furthermore, endurance cycling tests as shown in Fig. 3d reveal the robust switching endurance of the device. Under repeated program/erase cycling (±25 V, 0.1 s pulses), the memory window maintained a stable on/off ratio for over 1,000 cycles without signs of fatigue. The combination of multi-level programmability, long-term retention and excellent cycling endurance establishes this Te-based floating-gate device as a highly promising platform for reliable non-volatile photonic-electronic memory and in-memory computing systems.

**Ultrafast operation capability**

To assess the suitability of the device for high-speed photonic computing, we examined its transient response to nanosecond electrical pulses. Although static measurements confirm reliable switching, practical memories, particularly those targeted for electro-optical hybrid systems, require write speeds compatible with the bandwidth of modern logic and computing circuits. As shown in Fig. 4, ultrafast switching is achieved using voltage pulses with a full width at half maximum (FWHM) of 70 ns. This switching speed markedly exceeds that of electrically driven phase-change and thermo-optic memories, which are typically limited to microsecond–millisecond timescales by crystallization kinetics or thermal relaxation[22,45]. The nanosecond operation is consistent with intrinsic charge-tunnelling dynamics in high-performance van der Waals floating-gate devices[37,46], and uses its high electronic switching speed to set a non-volatile, optically readable state.

Specifically, the device shows precise, analog control over the stored charge density at a 70 ns programming speed. Applying a sequence of positive pulses (+27 V, 70 ns) resulted in a stepwise decrease in the MIR PL intensity (Fig. 4a), corresponding to the incremental depletion of holes. Conversely, a sequence of negative pulses (-18 V, 70 ns) produced a stepwise increase in the PL intensity (Fig. 4b), corresponding to the controlled injection of holes. Critically, the memory states can be deterministically tuned by modulating the pulse amplitude and duration, as the carrier injection density is highly sensitive to the applied electric field and temporal width (Supplementary Figs. 5 and 6). Remarkably, the device resolves over 16 distinct PL levels in both programming and erasing directions, confirming a storage capacity equivalent to 4 bits per cell. The switching speed is primarily constrained by the temporal resolution of the stimulus electrical pulse and the resolving capability of the PL levels, both of which are determined by the stimulation and detection equipment used in the measurement. While all the data presented in the Fig. 4 are stimulated with a 70-ns electrical pulse, a 50-ns pulse still stimulates an effective switch of the PL signal, as demonstrated in Fig. 5c and Supplementary Fig. 7 and 8. We note that the actual switching speed of LuminoMem could be even faster. This quasi-continuous modulation overcomes the density limitations of binary optical switches while preserving non-volatility.

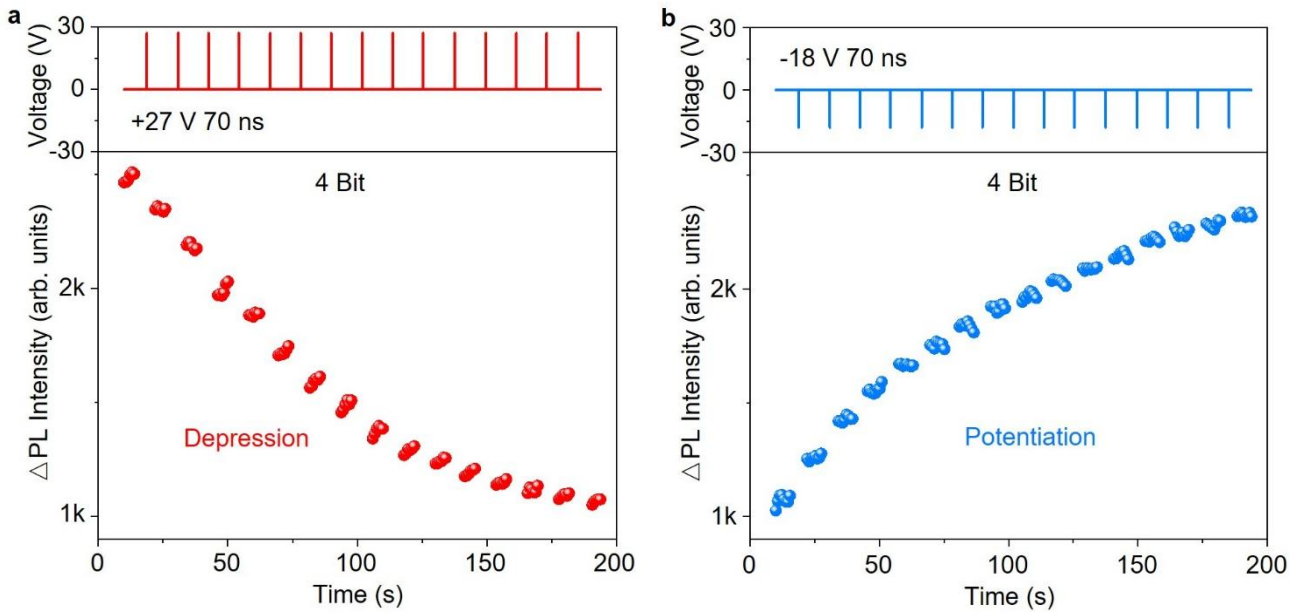

**Fig. 4 | Ultrafast switching dynamics and synaptic plasticity. a,** Stepwise decrease of PL intensity in response to a sequence of ultrafast pulses (amplitude: +27 V, FWHM: 70 ns), corresponding to the incremental depletion of holes and a gradual weakening of the synaptic weight. **b**, Stepwise increase of PL intensity under a sequence of ultrafast pulses (amplitude: -18 V, FWHM: 70 ns), reflecting controlled hole injection and a strengthening of the synaptic connection. These results demonstrate reproducible 4-bit storage per cell and high-speed, linear synaptic weight updates, which are essential for optical neuromorphic computing.

The nanosecond switching speed, which is rare even in state-of-the-art electronic flash memories, stems from the atomically sharp, defect-free van der Waals interfaces and strong electrical field across the h-BN layer[37,46]. These features collectively enable near-ballistic Fowler–Nordheim tunnelling. In conventional silicon floating-gate devices, dangling bonds and deep-level interface states at the semiconductor/dielectric boundary act as detrimental trapping centers. These defects lead to trap-assisted tunneling or charge-scattering, causing a significant lag in injection kinetics. By contrast, the pristine Gr/h-BN/Te heterostructure suppresses such parasitic pathways, achieving minimal electronic latency[47]. Furthermore, the high dielectric strength of few-layer h-BN further sustains large local electric fields without breakdown[46], and the microscale device footprint reduces parasitic RC delays. Consequently, the inherent properties of the heterostructure minimize the dominant non-ideal factors that plague conventional devices, thereby enabling an ultrafast switching speed.

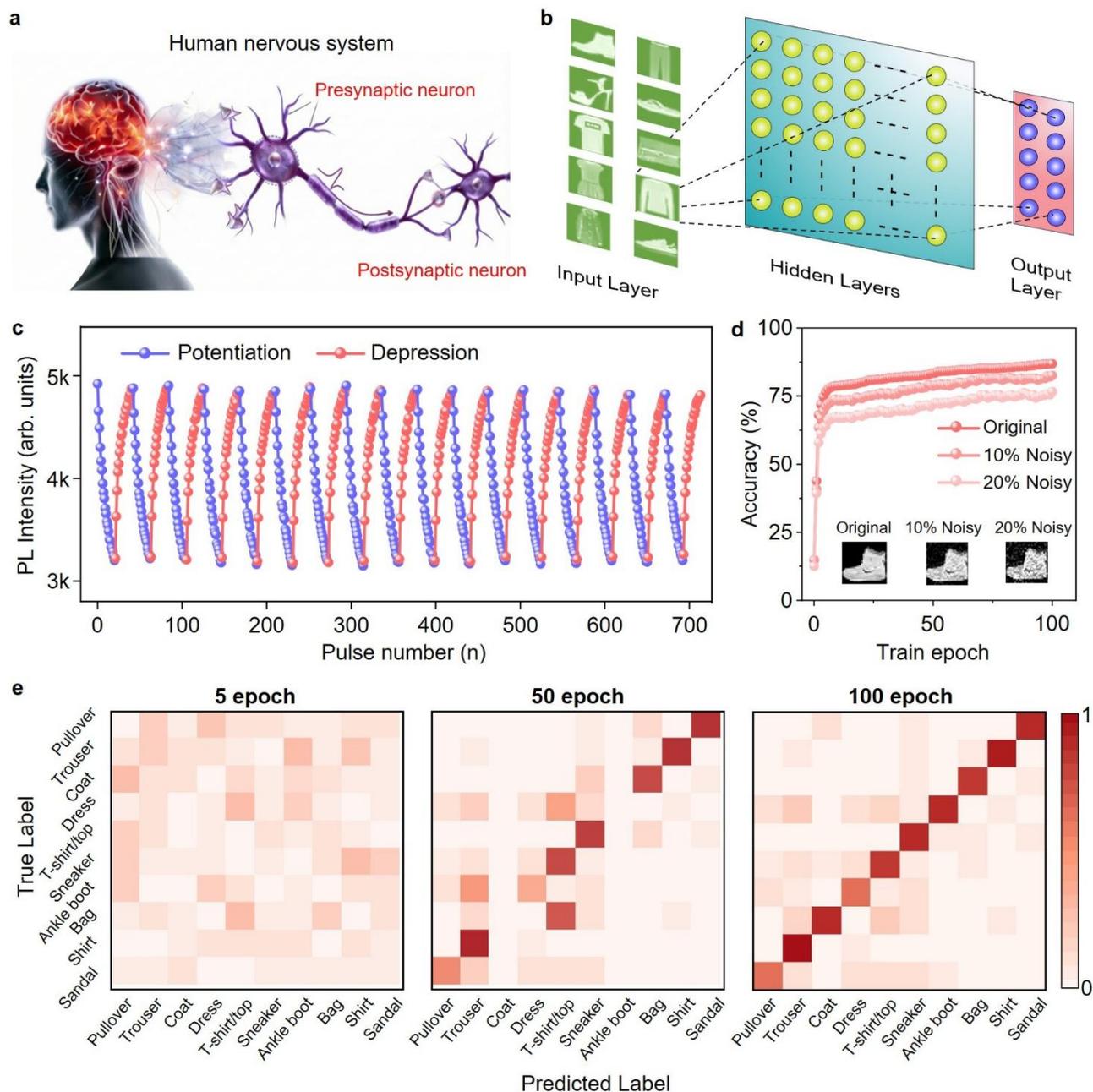

**Fig. 5 | Neuromorphic computing demonstration for fashion image recognition. a,** Schematic diagram of the human nervous system. **b,** Simulated three-layer ANN for classifying 10

categories from the Fashion-MNIST dataset. **c,** Seventeen potentiation and depression cycles of the device with 20 stable weight updates, emulating synaptic plasticity. **d,** Classification accuracy of the ANN under original, 10% and 20% noisy input conditions, demonstrating fault tolerance. **e,** Evolution of the confusion matrix (multicolor display) over training, where the strengthening diagonal (from initial 5 to 100 epochs) validates the network's successful learning and discriminative capability.

**Neural network inference based on LuminoMem device characteristics**

The linear, analogue modulation of LuminoMem PL demonstrated in Fig. 4 naturally emulates synaptic plasticity. In neuromorphic terms, stepwise increases in PL intensity correspond to long-term potentiation (LTP), whereas stepwise decreases emulate long-term depression (LTD). The ability to finely tune optical synaptic weight using ultrafast electrical pulses, achieving adjustable weight-update granularity through various pulse configurations (Supplementary Fig. 7), underscores LuminoMem's promise as a high-speed, linear synapse for infrared optical neural networks [48,49].

Inspired by the highly parallel and interconnected architecture of the human brain (Fig. 5a), we constructed a three-layer backpropagation artificial neural network (BP-ANN) using the LuminoMem array as synapses. We benchmarked the network's capability on the Fashion-MNIST dataset using the CrossSim simulator, mapping the measured potentiation-depression characteristics into the network model (Fig. 5b). The essential synaptic programmability of the device was meticulously optimized by modulating programming parameters, including pulse voltage, duration, and readout power (Supplementary Fig. 8). Under these optimized conditions (specifically +30 V with 300 ns for potentiation and −26 V with 50 ns for depression), the device exhibits excellent weight-update characteristics, as confirmed by 17 potentiation-depression cycles (Fig. 5c). These cycles demonstrate 20 highly stable and repeatable switching weights. While the weights are updated by ultrafast electrical pulses, the computation results are output as MIR PL signals, providing a pathway for direct optical processing in the MIR regime.

To evaluate the fault tolerance of this neural network, we introduced 10% and 20% random noise into the "ankle boot" category images. As shown in Fig. 5d, after 100 training epochs, the recognition accuracies for the pristine, 10%-noise, and 20%-noise datasets reached approximately 86.9%, 82.5%, and 76.6%, respectively. The sustained accuracy under substantial input perturbation highlights the network's robust learning capability. Consistently, the evolution of the confusion matrix during the training process (Fig. 5e) shows progressively enhanced diagonal dominance, indicating improved class separability and reliable classification across all categories.

**Conclusions**

The non-volatile LuminoMem helps close the long-standing gap between fast optical processing and comparatively slow, conversion-heavy electronic storage. By exploiting a vdW floating-gate architecture in which Weyl semiconductor Te simultaneously serves as charge storage and light-emitting medium, LuminoMem achieves 70-ns electrical programming with zero static power consumption. Data are accessed through direct optical readout in MIR, avoiding repeated electro–optic conversions and their associated latency and energy overheads. The device supports 4-bit multilevel storage with endurance exceeding 1,000 cycles, enabling analog weight representation. These device characteristics are further exploited for fashion image classification by constructing an artificial photonic synaptic array in a three-layer BP-ANN, thereby preliminarily demonstrating its potential for neural network application.

The ability to electrically program and update synaptic weights while performing high-speed optical readout positions LuminoMem as a promising building block for large-scale photonic neural networks. Integration with wavelength-division multiplexing, photonic interconnect, and CMOS-compatible control circuitry could enable massively parallel, in-memory computing architectures that surpass the bandwidth and energy efficiency limits of purely electronic systems. Beyond the performance demonstrated here, Te offers substantial additional opportunities. As a Weyl material, Te exhibits topologically enhanced nonlinear optical responses[50,51] and polarization-tunable emission[52], which could be exploited to improve device functionality and increase information transmission capacity in photonic neural networks. Furthermore, the present optoelectronic operating scheme is not fundamentally limiting: PL-based optical readout may be replaced by electrically driven electroluminescence, while electrical programming and erasing could be extended to fully optical operation through modest modifications of the device architecture. These advances may further broaden the applicability of LuminoMem in scalable photonic computing systems, while further advances in device uniformity, array-level integration, and on-chip learning schemes, may pave the way toward scalable photonic processors for neuromorphic computing, reconfigurable MIR optical communications and intelligent sensing.

## Methods

### Synthesis of Te nanoflakes and fabrication of LuminoMem devices

Te nanoflakes were grown through a hydrothermal method[38,53]. First, 3 g of polyvinylpyrrolidone (PVP, molecular weight = 58000) was dissolved in 32 mL of deionized (DI) water. Subsequently, 92 mg of $Na_2TeO_3$ was put into the PVP solution with continuous stirring. After that, 3.32 mL of ammonium hydroxide solution (25–28%, wt/wt%) and 1.68 mL of hydrazine hydrate (80%, wt/wt%) were added into the solution. Following 5 minutes of magnetic stirring, the solution was transferred into a 50 mL Teflon-lined stainless-steel autoclave and maintained at 180 °C for 10 hours. The resulting product was washed with deionized water to remove any residual ions.

To fabricate the memory devices, the synthesized Te nanoflakes were dispersed in ethanol and drop-cast onto a p$^{++}$-Si/SiO$_2$ (285 nm) substrate. Separately, a graphite/h-BN heterostructure was pre-assembled using a polycarbonate/polydimethylsiloxane (PC/PDMS) sequential pick-up technique at 80 °C. This heterostructure was then precisely aligned and laminated onto a selected Te nanoflake at 170 °C. Following the lamination, the sacrificial PC layer was dissolved in chloroform, and the sample was thoroughly rinsed with acetone and isopropanol to eliminate organic residues. This dry transfer process was employed to ensure the formation of atomically clean vdW interfaces between the Gr, h-BN and Te layers[54]. Electrical contacts and electrode geometries were defined using electron beam lithography (EBL). Finally, Ti/Pd/Au (0.5/20/70 nm) metal layers were deposited via electron-beam evaporation to form the source, drain, and gate contacts.

**PL readout with electrical gating**

PL readout was conducted using a self-built micro-PL measurement system. A steady 1550-nm CW laser was employed to excite the Te nanoflake. A ×40 reflective objective with a numerical aperture of 0.5 was used to focus the excitation laser onto the sample and collect the PL signal. All measurements were performed under vacuum in a liquid-helium-flow cryostat with electrical contact, enabling temperature control from 25 K. The cryostat was mounted on an XY scanner stage for spatial scanning. The collected PL signal was delivered to a spectrometer (Princeton SP2500i) equipped with a 150 g mm$^{-1}$ grating and detected by a liquid nitrogen-cooled single-channel InSb photodetector with a preamplifier. The detector output was processed with a lock-in amplifier referenced to an optical chopper that modulated the excitation laser at 521 Hz. For memory measurements, control-gate voltages were applied using Keithley 2612B source meters for 0.1 s pulses and a Keithley 4200 semiconductor characterization system (4200-SCS) for shorter pulses.

**Hardware-Aware neural network simulations**

A hardware-aware three-layer fully connected BP-ANN was implemented on the CrossSim platform for neuromorphic pattern recognition. The network adopted an input-hidden-output architecture with 784×64×10 neurons for the Fashion-MNIST dataset (28×28 grayscale images, 10 classes). Synaptic weights were mapped to the PL intensity, where weight potentiation and depression corresponded to the experimentally observed LTP and LTD characteristics. Input noise of 0–20% amplitude was added to assess robustness, and classification performance was evaluated via accuracy and confusion matrices after training.

To capture hardware non-idealities, a lookup table (LUT) describing the measured PL-based weight-update behavior was embedded into both the forward computation and weight update within CrossSim, forming a hardware-aware training loop. Quantization and range constraints were included to emulate the limited precision and dynamic range of peripheral circuits. Network parameters were trained using standard backpropagation with gradient descent and cross-entropy loss, and accuracy was tested on an independent validation set.

## Data availability

The data that support the findings of this study are available from the corresponding author upon reasonable request. Source data are provided with this paper.

**Acknowledgements**

This project was supported by the National Natural Science Foundation of China (Grant Nos. 62325401 [D.S.] and 62431018 [J.L.]) and the National Key Research and Development Program of China (Grant No. 2021YFA1400100 [D.S.], 2022YFA1204300 [A.L.P.], 2023YFA1406300 [C.G.Z.] and 2024YFA1208400 [J.L.]).  The authors would also like to acknowledge the National Natural Science Foundation of China (Grant Nos. 12034001 [D.S.], 52221001 [A.L.P.] and 62227822 [D.S.]), the Key Program of Science and Technology Department of Hunan Province (2019XK2001[A.L.P.] and 2020XK2001[A.L.P.]), and the Key Research and Development Plan of Hunan Province (2023GK2012[A.L.P.]).

**Author contributions**

J.L. and D.S. conceived the idea and supervised the project. B.C. synthesized the Te nanoflakes under the supervision of C.G.Z. and L.L.; D.L.L. performed the measurements and basic characterizations with the help from S.Y.W., Y.W., D.L., Y.C.C., M.Y.Q., D.H.Y., and J.S. under the supervision of A.L.P. and D.S.; Y.W. performed the ANN simulation under the supervision of J.L.; D.L.L., J.L. and D.S. wrote the manuscript with input and feedback from all of the authors.

**Competing interests**

The authors declare no competing interests.